\documentclass[reprint,showpacs,preprintnumbers,amsmath,amssymb]{revtex4-1}
\pdfoutput=1
\usepackage[pdftex]{graphicx}
\usepackage[unicode=true, bookmarks=true,bookmarksnumbered=true,bookmarksopen=true, breaklinks=false,backref=false,pagebackref=false, colorlinks=true]{hyperref}
\hypersetup{pdftitle={Colour fields for the static gluon-quark-antiquark system,
and a microscopic study of the Casimir scaling},
pdfauthor={Marco Cardoso, Nuno Cardoso, Pedro Bicudo},
linkcolor=black,citecolor=black, urlcolor=black, filecolor=black,pdfpagelayout=OneColumn,pdfnewwindow=true,pdfstartview=XYZ, plainpages=false}
\usepackage{color}
\usepackage{amsmath}
\usepackage{braket}
\usepackage[T1]{fontenc}
\usepackage{url}
\usepackage[english]{babel}
\usepackage{graphicx}
\usepackage{dcolumn}
\usepackage{bm}
\usepackage{subfig}
\usepackage{float}
\usepackage{url}
\usepackage{braket}
\newcommand{\Tr}{\mbox{\rm Tr}}
\newcommand{\ReC}{\mbox{\rm Re}}
\usepackage{verbatim}

\begin{document}

\title{
Lattice QCD computation of the colour fields for the static hybrid
\\
quark-gluon-antiquark system, and microscopic study of the Casimir scaling}
\author{
M. Cardoso}
\author{
N. Cardoso}
\author{
P. Bicudo}
\affiliation{CFTP, Departamento de F\'{\i}sica, Instituto Superior T\'{e}cnico,
Av. Rovisco Pais, 1049-001 Lisboa, Portugal}

\begin{abstract}
The chromoelectric and chromomagnetic fields, created by a static gluon-quark-antiquark system, are computed in quenched SU(3) lattice QCD, in a $24^3\times 48$ lattice at $\beta=6.2$ and $a=0.07261(85)\,fm$. We compute the hybrid Wilson Loop with two spatial geometries, one with a U shape and another with an L shape. The particular cases of the two gluon glueball and quark-antiquark are also studied, and the Casimir scaling is investigated in a microscopic perspective. This microscopic study of the colour fields is relevant to understand the structure of hadrons, in particular of the hybrid excitation of mesons. This also contributes to understand confinement with flux tubes and to discriminate between the models of fundamental versus adjoint confining strings, analogous to type-II and type-I superconductivity.
\end{abstract}
\maketitle

\section{Introduction}

Here we present the first Lattice QCD study of the chromoelectric and chromomagnetic fields, created by a static gluon-quark-antiquark system.
Although the colour fields have been extensively studied for the quark-antiquark,
\cite{Haymaker:1994fm,Barczyk:1994yw,Trottier:1995if,Ichie:2002dy},
for three quarks
\cite{Ichie:2002dy,Okiharu:2003vt,Suganuma:2004hu,Takahashi:2004kc,Signal:2008zn},
for the hybrid only the static potential has been studied so far
\cite{Cardoso:2007dc,Bicudo:2007xp}.

\begin{figure}[h!]
\begin{center}
    \includegraphics[width=7cm]{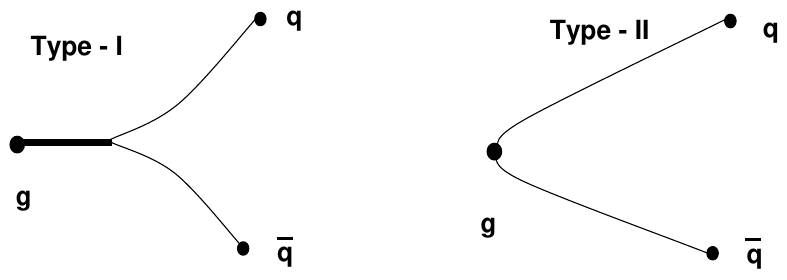}
    \caption{String attraction and fusion, and string repulsion, respectively in type I and II superconductors.}
    \label{superconductor}
\end{center}
\end{figure}

The hybrid static potential is also relevant to understand the nature of confinement and of Casimir Scaling, since with the hybrid potential we can interpolate between the gluon-gluon interaction and the quark-antiquark interactions which are particular cases of the hybrid static potential.
The first study of the static gluon-gluon interaction was performed by Michael \cite{Michael:1985ne,Campbell:1985kp}, and Bali \cite{Bali:2000un} extended this study to other SU(3) representations, leading to the Casimir Scaling picture.
Bicudo et al. \cite{Bicudo:2007xp} and Cardoso et al. \cite{Cardoso:2007dc} studied the static gluon-quark-antiquark potential and showed that when the segments gluon-quark and gluon-antiquark are perpendicular, the potential $V$ is compatible with the confinement realized with a pair of fundamental strings, one linking the gluon to the quark and the other linking the same gluon to the antiquark.
For parallel and superposed segments, however, the total string tension becomes larger and is in agreement with the Casimir Scaling measured by Bali \cite{Bali:2000un}.
Bicudo, Cardoso and Oliveira established an analogy between the static potential and a type-II superconductor for the confinement in QCD, illustrated in Fig. \ref{superconductor}, with repulsion of the fundamental strings and with the string tension of the first topological excitation of the string (the adjoint string) larger than the double of the fundamental string tension. In type-I superconductor the fundamental strings would be attracted and would fuse into an adjoint string. With the computation of the flux tubes we can further understand, microscopically the Casimir Scaling. For instance Semay
\cite{Semay:2004br}
presented a model for Casimir Scaling, based on a shape of flux tubes independent of the colour SU(3) representation, and we can test it.

In this paper, we investigate the chromoelectric and chromomagnetic fields, and the resulting lagrangian and energy density distributions around a static gluon-quark-antiquark system in quenched SU(3) lattice QCD.
In section II, we introduce the lattice QCD formulation. We briefly review the Wilson loop for this system, which was used in Bicudo et al. \cite{Bicudo:2007xp} and Cardoso et al. \cite{Cardoso:2007dc}, and show how we compute the colour fields and the lagrangian and energy density distribution.
In section III, the numerical results are shown, including several density plots of the chromo flux tubes, and longitudinal plots of the chromo field profiles.
Finally, we present the conclusion in section IV.

\section{The Wilson Loops and Colour Fields}
\begin{figure}[h]
\begin{centering}
    \subfloat[\label{loop0}]{
\begin{centering}
    \includegraphics[width=4cm]{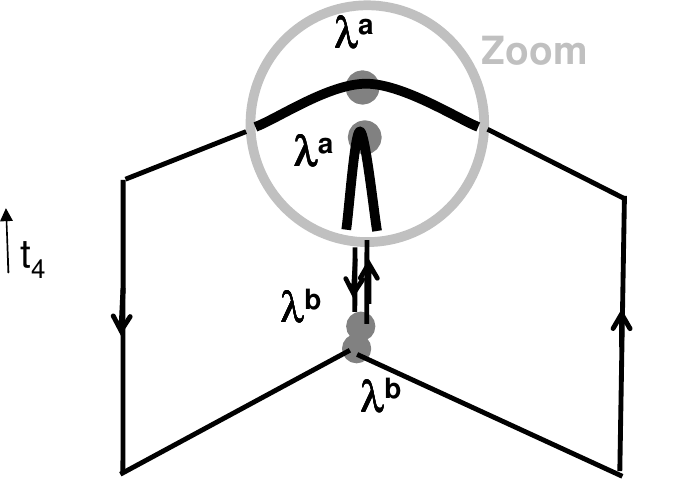}
\par\end{centering}}
\\
    \subfloat[\label{loop1}]{
\begin{centering}
    \includegraphics[width=6cm]{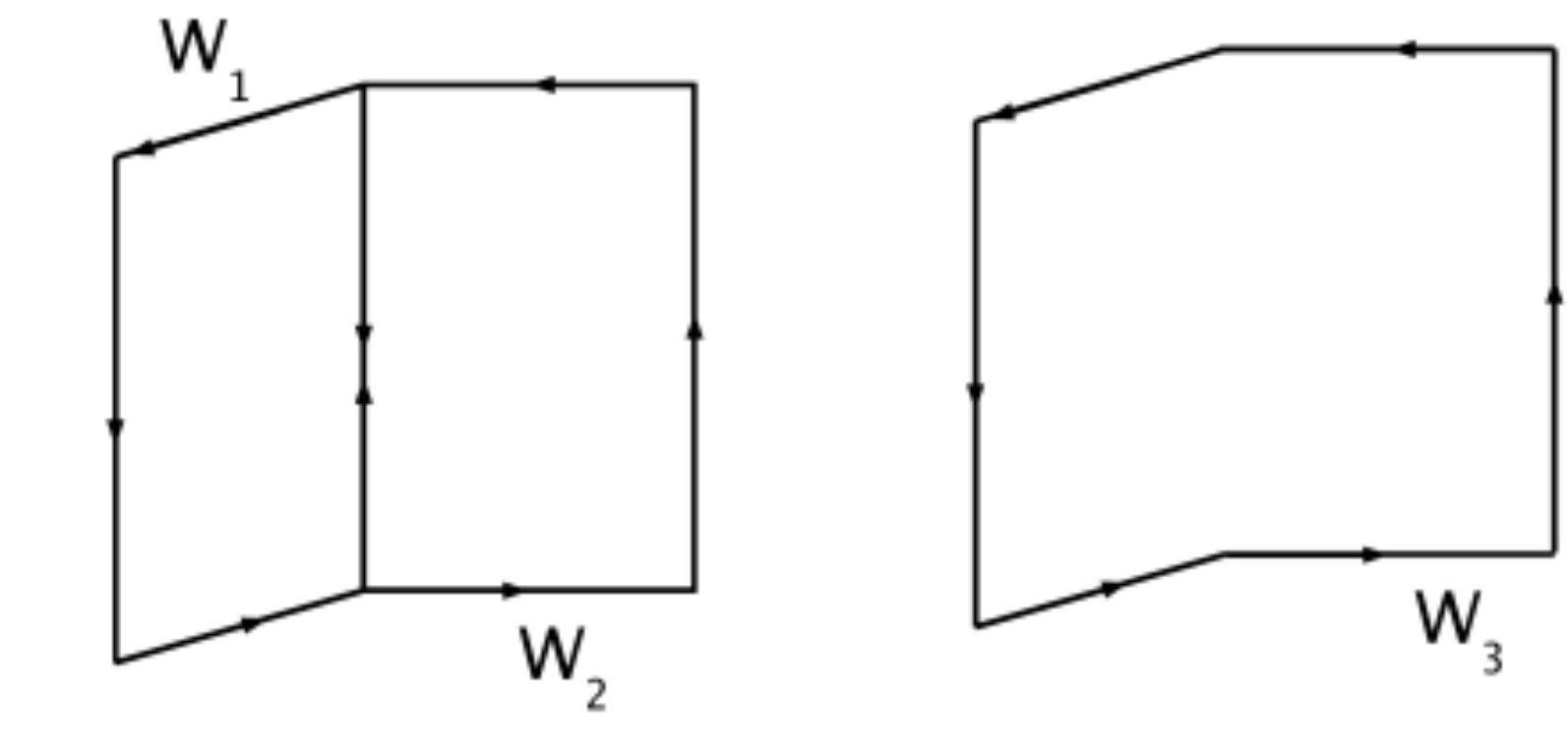}
\par\end{centering}}
\par\end{centering}
    \caption{\protect\subref{loop0} Wilson loop for the $gq\overline{q}$ and equivalent position of the static antiquark, gluon, and quark. \protect\subref{loop1} Simple Wilson loops that make the $gq\overline{q}$ Wilson loop.}
    \label{loop}
\end{figure}

In principle, any Wilson loop with a geometry similar to that represented in Fig. \ref{loop0}, describing correctly the quantum numbers of the hybrid, is appropriate, although the signal to noise ratio may depend on the choice of the Wilson loop.
A correct Wilson loop must include an SU(3) octet (the gluon), an SU(3) triplet (the quark) and an SU(3) antitriplet (the antiquark), as well as the connection between the three links of the gluon, the quark and the antiquark.

We construct the gluon-quark-antiquark Wilson loop from the two-color-octet meson operator,
\begin{equation}
    \mathcal{O}(x)=\frac{1}{4}\left[\overline{q}(x)\lambda^a\Gamma_1 q(x)\right]\left[\overline{q}(x)\lambda^a\Gamma_2 q(x)\right]
    \label{wlmop}
\end{equation}
where $\lambda^a$ are the Gellmann SU(3) colour matrices, and where $\Gamma_i$ are spinor matrices. Using the lattice links to comply with gauge invariance, the second operator in Eq. (\ref{wlmop}) can be made nonlocal to separate the quark and the antiquark from the gluon,
\begin{eqnarray}
    \mathcal{O}(x) & = & \frac{1}{4}\Big[\overline{q}(x)\lambda^a\Gamma_1q(x)\Big]\nonumber \\
    & & \Big[\overline{q}\left(x-x_1\hat{\mu}_1\right)U_{\mu_1}\left(x-x_1\hat{\mu}_1\right)\cdots U_{\mu_1}\left(x-\hat{\mu}_1\right)\nonumber \\
    & & \lambda^a\Gamma_2 U_{\mu_2}\left(x\right)\cdots U_{\mu_2}\left(x+\left(x_2-1\right)\hat{\mu}_2\right)\nonumber \\
    & & q\left(x+x_2\hat{\mu}_2\right)\Big]\ .
    \label{wlmop1}
\end{eqnarray}
The contraction of the quark field operators, assuming that all quarks are of different nature, gives rise to the gluon operator,
\begin{eqnarray}
W_{gq\overline{q}} & = & \frac{1}{16}\Tr \Big[
  U^\dagger_4 (t-1,x) \cdots U^\dagger_4 (0,x)  ~  \lambda^b
\nonumber \\
    & & U_4 (0,x) \cdots U_4 (t-1,x)\lambda^a \Big]\nonumber \\
    & &\Tr\Big[U_{\mu_2} (t,x)\cdots U_{\mu_2}(t,x+(x_2-1)\hat{\mu}_2)\nonumber \\
    & &U^\dagger_4 (t-1,x + x_2 \hat{\mu}_2)\cdots U^\dagger_4 (0,x+x_2\hat{\mu}_2)\nonumber \\
    & &U^\dagger_{\mu_2} (0,x + (x_2-1) \hat{\mu}_2)\cdots U^\dagger_{\mu_2} (0,x) \lambda^b \nonumber \\
    & &U^\dagger_{\mu_1} (0,x - \hat{\mu}_1)\cdots
    U^\dagger_{\mu_1} (0,x - x_1 \hat{\mu}_1 )
    \nonumber \\
    & &U_4 (0,x - x_1 \hat{\mu}_1) \cdots
    U_4 (t-1,x-x_1\hat{\mu}_1)\nonumber \\
    & &U_{\mu_1} (t,x - x_1 \hat{\mu}_1) \cdots
    U_{\mu_1} (t,x - \hat{\mu}_1 ) \lambda^a \Big]\ .
\label{glue88}
\end{eqnarray}
Using the Fiertz relation,
\begin{equation}
\sum_a \, \left( \frac{\lambda^a}{2} \right)_{ij} \,
          \left( \frac{\lambda^a}{2} \right)_{kl} ~ = ~
    \frac{1}{2} \delta_{il} \delta_{jk} -
     \frac{1}{6} \delta_{ij} \delta_{kl}
\label{exchange+identity}
\end{equation}
we can prove that
\begin{equation}
    W_{gq\overline{q}}=W_1 W_2 - \frac{1}{3}W_3
\label{wloopqqg}
\end{equation}
where $W_1$, $W_2$ and $W_3$ are the simple Wilson loops shown in Fig. \ref{loop1}.
Importantly for the study of the Casimir Scaling,
when $r_1=0$, $W_1=3$ and $W_2=W_3$, the operator reduces to the mesonic Wilson loop
and when $\mu=\nu$ and $r_1=r_2=r$, $W_2=W_1^{\dagger}$ and $W_3=3$, $W_{gq\overline{q}}$ reduces
to $W_{gq\overline{q}}(r,r,t)=|W(r,t)|^2-1$, that is the Wilson loop in the adjoint representation
used to compute the potential between two static gluons.

\begin{figure}
\begin{centering}
    \subfloat[U shape geometry.\label{fig:shapeU}]{
\begin{centering}
    \includegraphics[height=3cm]{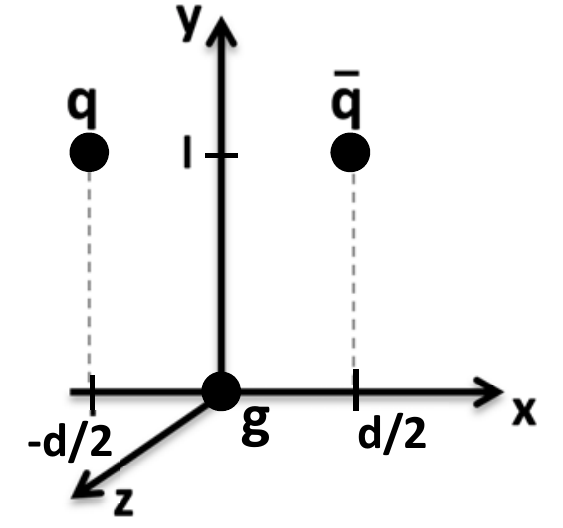}
\par\end{centering}}
    \subfloat[L shape geometry.\label{fig:shapeL}]{
\begin{centering}
    \includegraphics[height=3cm]{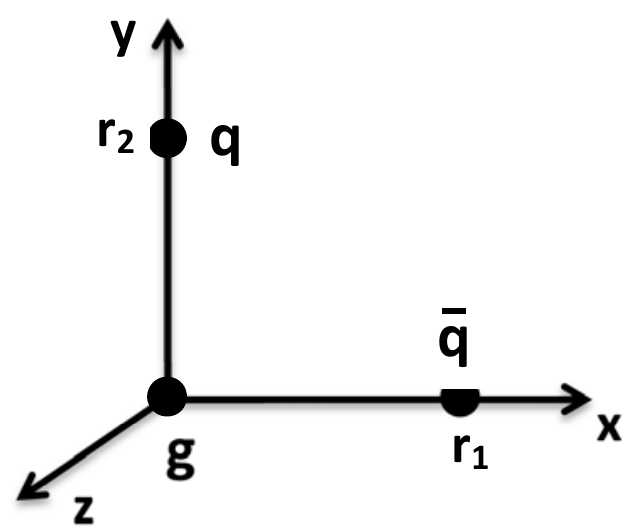}
\par\end{centering}}
\par\end{centering}
    \caption{gluon-quark-antiquark geometries, U and L shapes.}
    \label{shape}
\end{figure}

In order to improve the signal to noise ratio of the Wilson loop, the links are replaced by "fat links",
\begin{eqnarray}
    U_{\mu}\left(s\right) & \rightarrow & P_{SU(3)}\frac{1}{1+6w}\Big(U_{\mu}\left(s\right)\nonumber \\
    & & + w \sum_{\mu\neq\nu}U_{\nu}\left(s\right) U_{\mu}\left(s+\nu\right)U_{\nu}^{\dagger}\left(s+\mu\right)\Big)\ .
\end{eqnarray}
We use $w = 0.2$ and iterate this procedure 25 times in the spatial direction.

We obtain the chromoelectric and chromomagnetic fields on the lattice, by using,
\begin{equation}
    \Braket{E^2_i}= \Braket{P_{0i}}-\frac{\Braket{W\,P_{0i}}}{\Braket{W}}
\end{equation}
and,
\begin{equation}
    \Braket{B^2_i}= \frac{\Braket{W\,P_{jk}}}{\Braket{W}}-\Braket{P_{jk}}
\end{equation}
where the $jk$ indices of the plaquette complement the index $i$ of the magnetic field,
and where the plaquette is given by
\begin{equation}
P_{\mu\nu}\left(s\right)=1 - \frac{1}{3} \ReC\,\Tr\left[ U_{\mu}(s) U_{\nu}(s+\mu) U_{\mu}^\dagger(s+\nu) U_{\nu}^\dagger(s) \right]\ .
\end{equation}
The energy ($\mathcal{H}$) and lagrangian ($\mathcal{L}$) densities are given by
\begin{equation}
    \mathcal{H} = \frac{1}{2}\left( \Braket{E^2} + \Braket{B^2}\right)\ ,
    \label{energy_density}
\end{equation}
\begin{equation}
    \mathcal{L} = \frac{1}{2}\left( \Braket{E^2} - \Braket{B^2}\right)\ .
    \label{lagrangian_density}
\end{equation}
Notice that we only apply the smearing technique to the Wilson loop.

\begin{figure*}
\begin{centering}
    \subfloat[$\Braket{E^2}$\label{fig:cfield_qqg_U_d_0_l_8_E}]{
\begin{centering}
    \includegraphics[width=3.4cm]{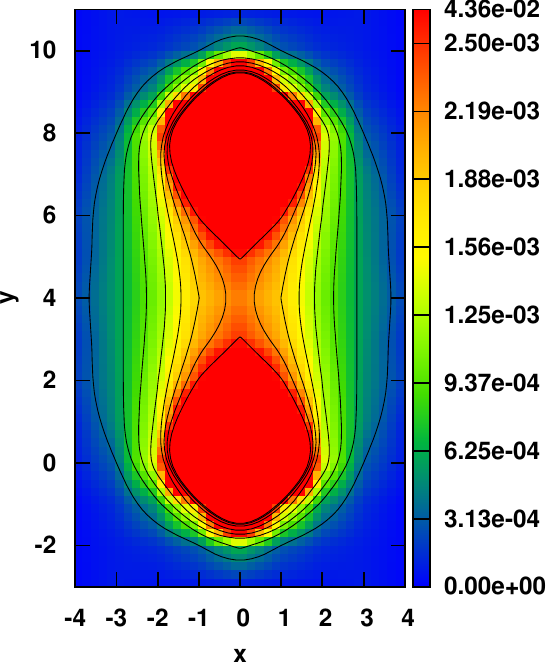}
\par\end{centering}}
    \subfloat[$-\Braket{B^2}$\label{fig:cfield_qqg_U_d_0_l_8_B}]{
\begin{centering}
    \includegraphics[width=3.4cm]{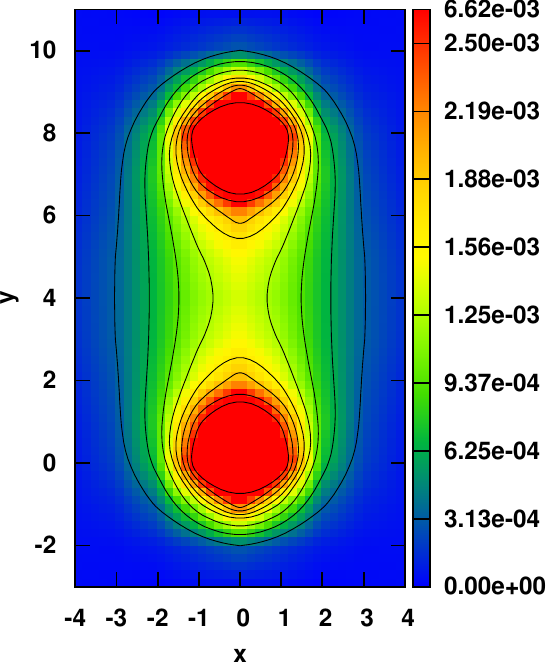}
\par\end{centering}}
    \subfloat[Energy Density\label{fig:cfield_qqg_U_d_0_l_8_Energ}]{
\begin{centering}
    \includegraphics[width=3.4cm]{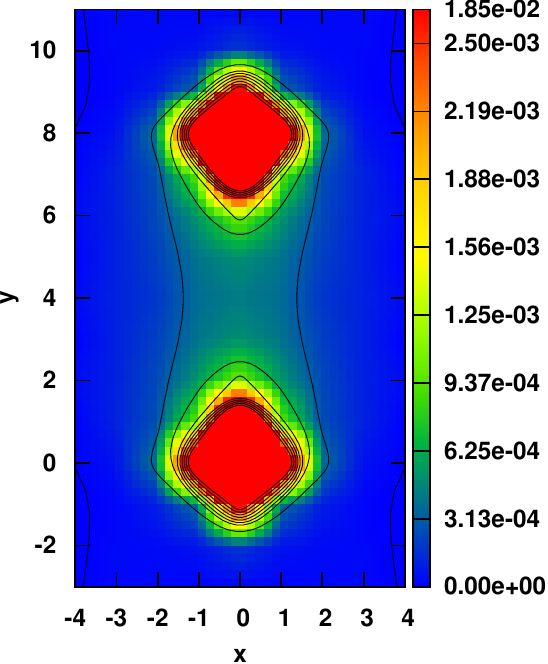}
\par\end{centering}}
    \subfloat[Lagrangian Density\label{fig:cfield_qqg_U_d_0_l_8_Act}]{
\begin{centering}
    \includegraphics[width=3.4cm]{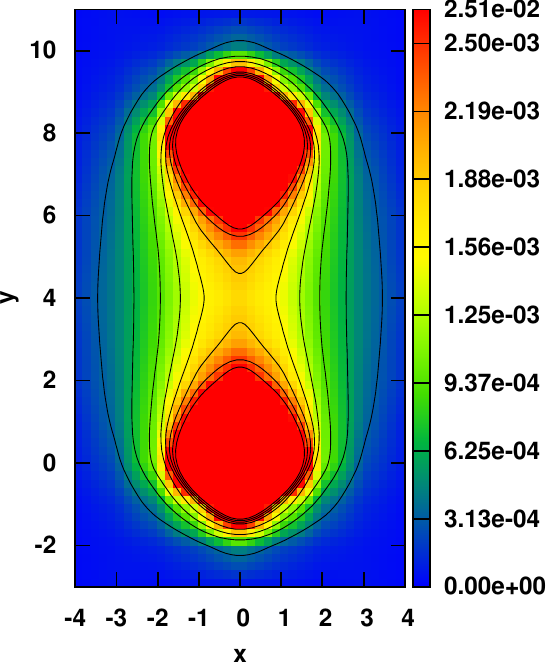}
\par\end{centering}}
\par\end{centering}
    \caption{Results for the static two gluon glueball. The energy density plot, \protect\subref{fig:cfield_qqg_U_d_0_l_8_Energ}, have a flux tube, responsible for the string tension, but we choose to put all the colour scales on the same value in order to be able to make a comparison with the values of the different fields, thus the flux tube energy is less visible. The top value of the colour scale is the maximum value of the field. The results are in lattice spacing units.}
    \label{glueball_field}
\end{figure*}

\begin{figure*}
\begin{centering}
    \subfloat[$\Braket{E^2}$\label{fig:cfield_qqg_L_r1_0_r2_8_E}]{
\begin{centering}
    \includegraphics[width=3.4cm]{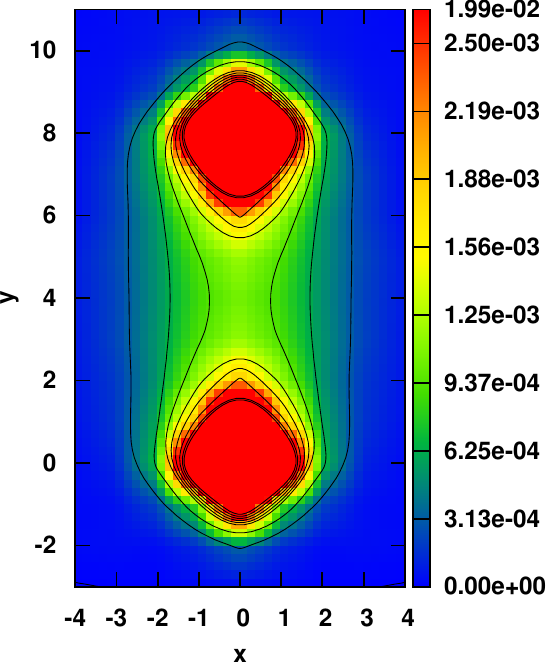}
\par\end{centering}}
    \subfloat[$-\Braket{B^2}$\label{fig:cfield_qqg_L_r1_0_r2_8_B}]{
\begin{centering}
    \includegraphics[width=3.4cm]{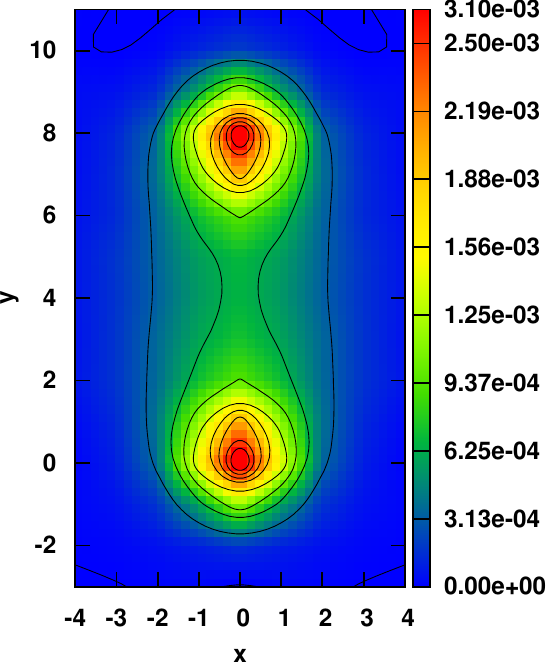}
\par\end{centering}}
    \subfloat[Energy Density\label{fig:cfield_qqg_L_r1_0_r2_8_Energ}]{
\begin{centering}
    \includegraphics[width=3.4cm]{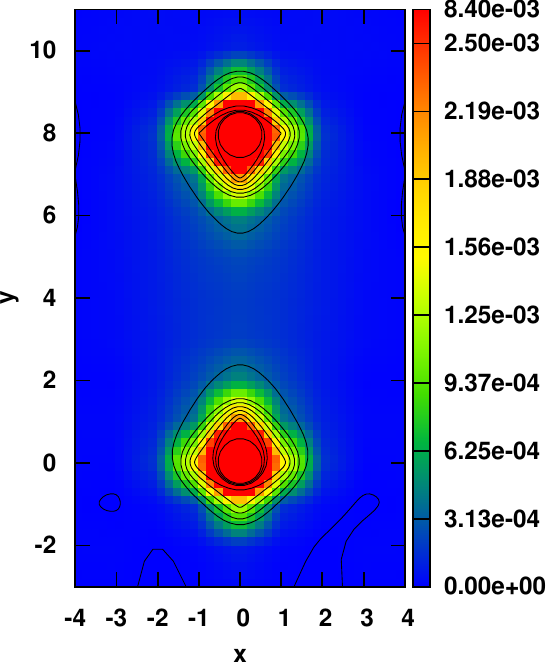}
\par\end{centering}}
    \subfloat[Lagrangian Density\label{fig:cfield_qqg_L_r1_0_r2_8_Act}]{
\begin{centering}
    \includegraphics[width=3.4cm]{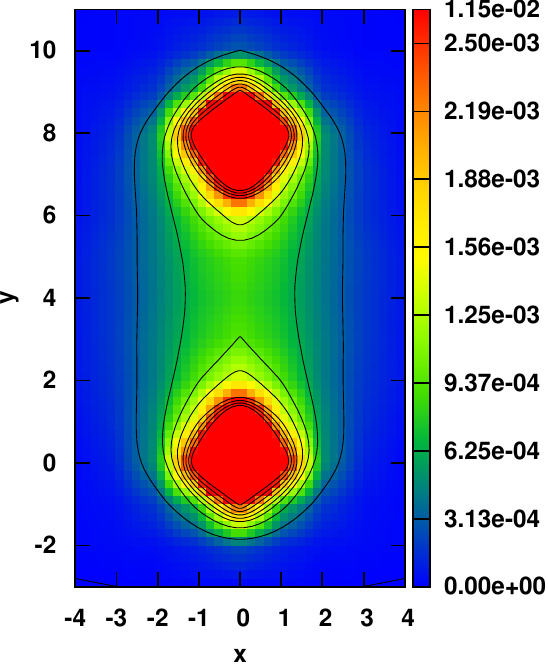}
\par\end{centering}}
\par\end{centering}
    \caption{Results for the static quark-antiquark system. The energy density plot, \protect\subref{fig:cfield_qqg_L_r1_0_r2_8_Energ}, have a flux tube, responsible for the string tension, but we choose to put all the colour scales on the same value in order to be able to make a comparison with the values of the different fields, thus the flux tube energy is less visible. The top value of the colour scale is the maximum value of the field. The results are in lattice spacing units.}
    \label{meson_field}
\end{figure*}

\section{Results}

Here we present the results of our simulations with 266 SU(3) configurations in a $24^3 \times 48$, $\beta = 6.2$ lattice, generated with the version 6 of the MILC code \cite{MILC}, via a combination of Cabbibo-Mariani and overrelaxed updates.
The results are presented in lattice spacing units, defined in Eq. (\ref{glue88}).

In this work two geometries for the hybrid system, gluon-quark-antiquark, are investigated: a U shape and a L shape geometry,
both defined in Fig. \ref{shape}.

In the U shape geometry, we only change the distance between quark and antiquark, $d=0,2,4,6$, and we fix the distance between gluon and quark-antiquark at $l=8$.
When the quark and the antiquark are superposed, $d = 0$, the system corresponds to a two gluon glueball, Fig. \ref{glueball_field}.

In the L shape geometry, we fix the distance between the gluon and quark at $r_2=8$ and the distance between the gluon and antiquark is changed,  $r_1=0,2,4,6,8$.
When the gluon and the antiquark are superposed, $r_1=0$, the system is equivalent to a meson. The results for the
meson system are presented in Fig. \ref{meson_field}.

\begin{figure}
\begin{center}
    \includegraphics[width=5.250cm]{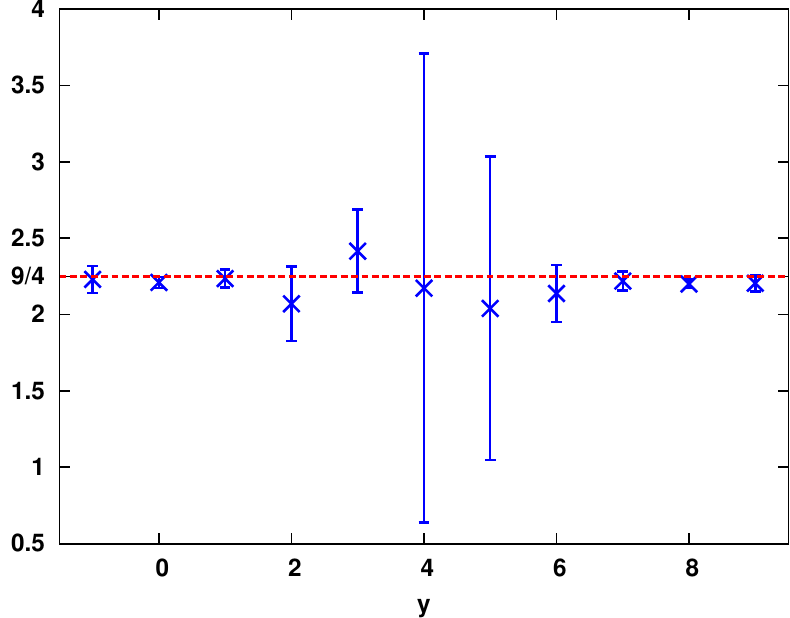}
    \caption{Results for the glueball ($d=0$ and $l=8$, U geometry) energy density over the meson ($r_1=0$ and $r_2=8$, L geometry) energy density for x=0. Casimir scaling, were the Casimir invariant $\lambda_i \cdot \lambda_j$ produces a factor of $9/4$ (broken line).}
    \label{glueball_meson_Energ}
\end{center}
\end{figure}

\begin{figure*}
\begin{centering}
    \subfloat[\label{fig:cfield_qqg_L_r1_8_r2_8_Ex}$\Braket{E_x^2}$]{
\begin{centering}
    \includegraphics[height=3.5cm]{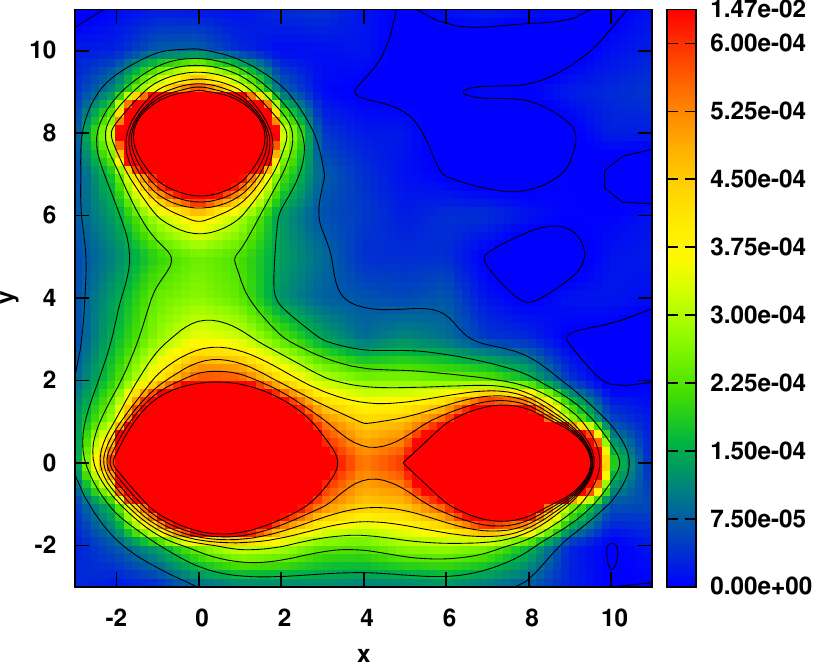}
\par\end{centering}}
    \subfloat[\label{fig:cfield_qqg_L_r1_8_r2_8_Ey}$\Braket{E_y^2}$]{
\begin{centering}
    \includegraphics[height=3.5cm]{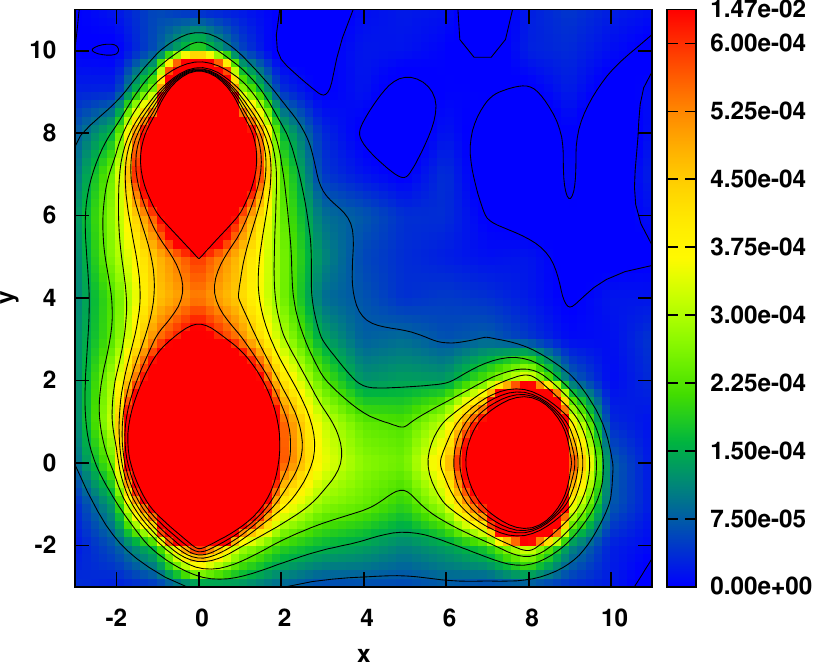}
\par\end{centering}}
    \subfloat[\label{fig:cfield_qqg_L_r1_8_r2_8_Ez}$\Braket{E_z^2}$]{
\begin{centering}
    \includegraphics[height=3.5cm]{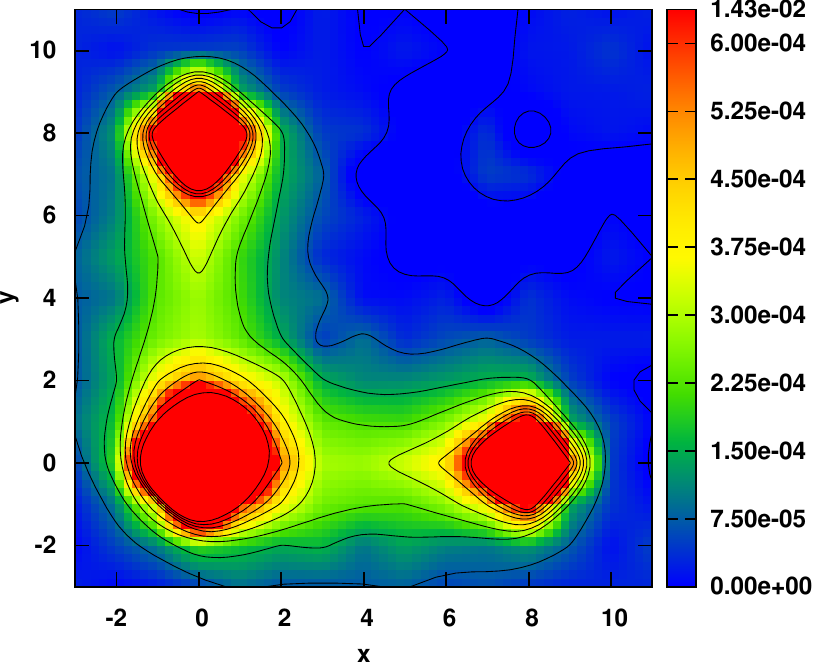}
\par\end{centering}}
    \subfloat[Energy Density\label{fig:cfield_qqg_L_r1_8_r2_8_Energ}]{
\begin{centering}
    \includegraphics[height=3.5cm]{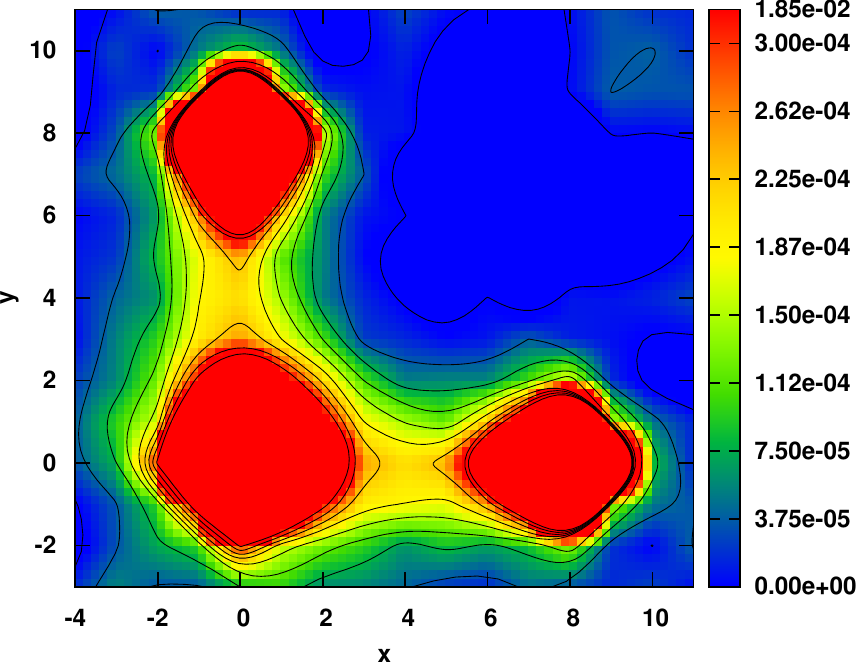}
\par\end{centering}}

    \subfloat[\label{fig:cfield_qqg_L_r1_8_r2_8_Bx}$-\Braket{B_x^2}$]{
\begin{centering}
    \includegraphics[height=3.5cm]{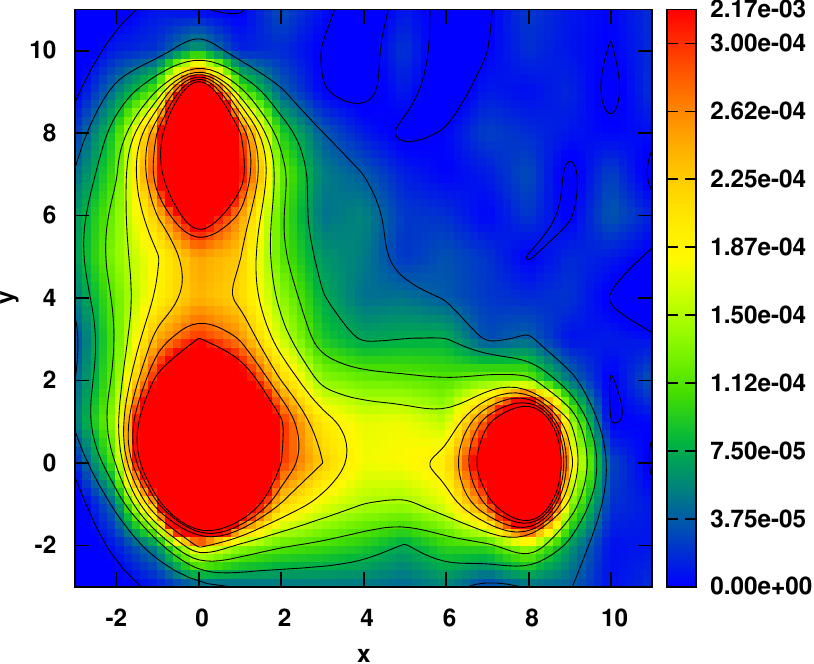}
\par\end{centering}}
    \subfloat[\label{fig:cfield_qqg_L_r1_8_r2_8_By}$-\Braket{B_y^2}$]{
\begin{centering}
    \includegraphics[height=3.5cm]{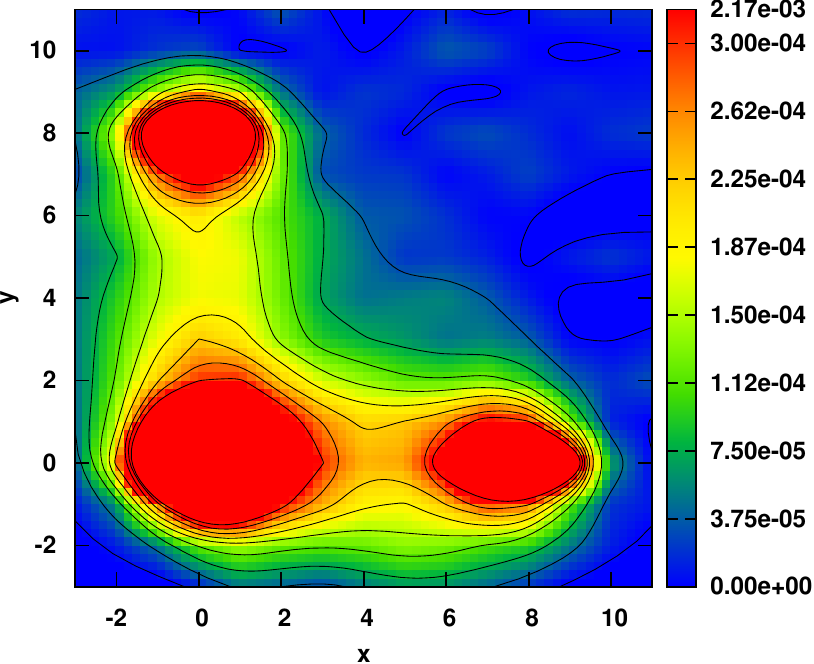}
\par\end{centering}}
    \subfloat[\label{fig:cfield_qqg_L_r1_8_r2_8_Bz}$-\Braket{B_z^2}$]{
\begin{centering}
    \includegraphics[height=3.5cm]{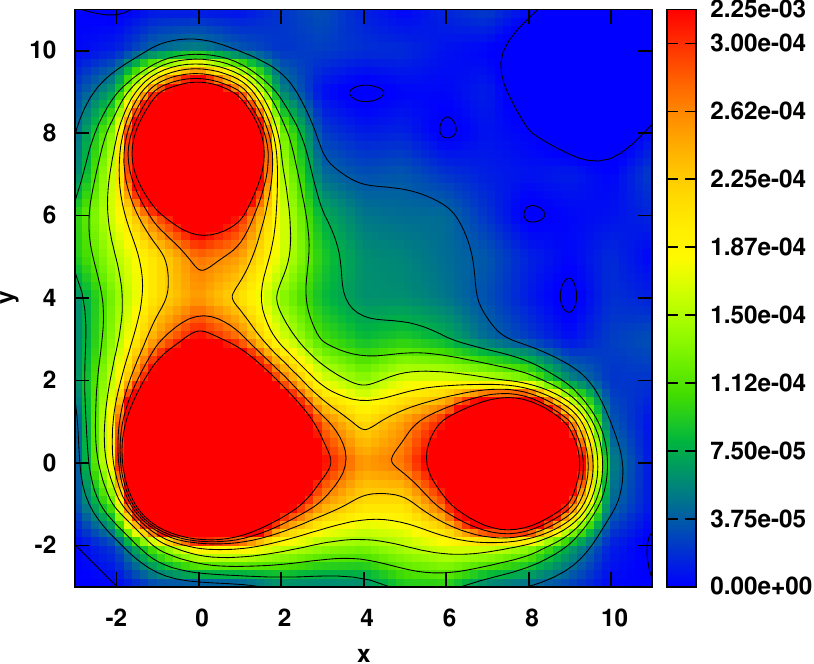}
\par\end{centering}}
    \subfloat[Lagrangian Density\label{fig:cfield_qqg_L_r1_8_r2_8_Act}]{
\begin{centering}
    \includegraphics[height=3.5cm]{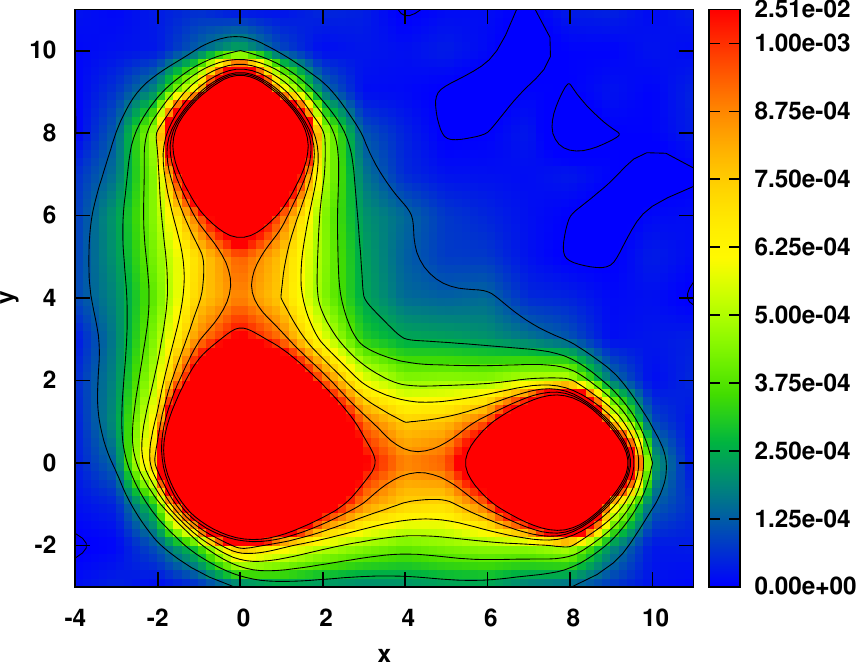}
\par\end{centering}}
\par\end{centering}
    \caption{Chromoelectric and chromomagnetic components and energy and lagrangian densities in the L shape geometry for $r_1=8$ and $r_2=8$. We use different colour scales to have a better view of the flux tube and the top value of the scale is the maximum value of the field. The results are in lattice spacing units.}
    \label{cfield_qqg_L_r1_8_r2_8_Ei_Bi}
\end{figure*}

\subsection{Flux Tube and Casimir Scaling}

First we discuss the results for the two degenerate cases, in which the system colapses into a two body system -
the meson (L geometry with $r_1 = 0$) and the two gluon glueball (U geometry with $d = 0$).
In the meson case we confirm the results obtained in previous works (for example \cite{Barczyk:1994yw}).
Not only in the meson case, but also in general, we have $\braket{E_{\parallel}^2} \geq \braket{E_{\perp}^2}
\geq | \braket{B_{\perp}^2} | \geq | \braket{B_{\parallel}^2} |$.
We also observe that $\braket{E^2} > 0$ and $\braket{B^2} < 0$, for all the studied geometries.
Since the absolute value of the chromoelectric field dominates
over the absolute value of the chromomagnetic field, there is a cancelation in energy density,
Eq. (\ref{energy_density}), and an enhancement in lagrangian density, Eq. (\ref{lagrangian_density}).

We measure the quotient between the energy densities of the meson system and of the glueball system,
in the mediatrix plane between the two particles ($x = 0$).
The results are shown in Fig. \ref{glueball_meson_Energ}. As can be seen, these results are consistent with
Casimir scaling, with a factor of $9/4$ between the energy density in the glueball and in the meson. This corresponds
to the formation of an adjoint string between the two gluons. The results are compatible with an identical
shape of the two flux tubes, but with a different density, and in this sense this agrees with the simple picture for
the Casimir Scaling of Semay,
\cite{Semay:2004br}.

The results for the fields in the case of the two gluon-glueball are given in Fig. \ref{glueball_field} and the meson in Fig. \ref{meson_field}.

\begin{figure*}
\begin{centering}
    \subfloat[U geometry at $y=4$.\label{fig:action_U_g_qaq}]{
\begin{centering}
    \includegraphics[height=6cm]{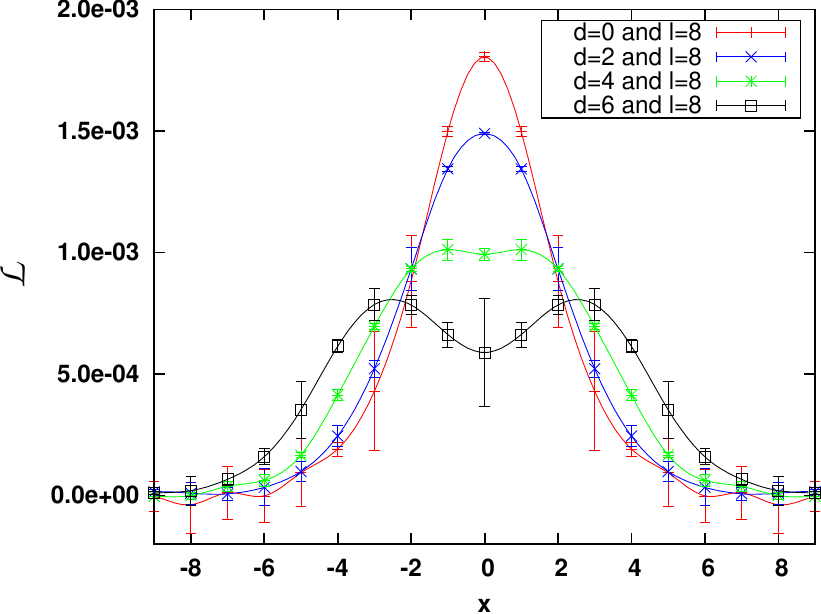}
\par\end{centering}}
    \subfloat[U geometry at $x=0$.\label{fig:action_U_gqq_x0}]{
\begin{centering}
    \includegraphics[height=6cm]{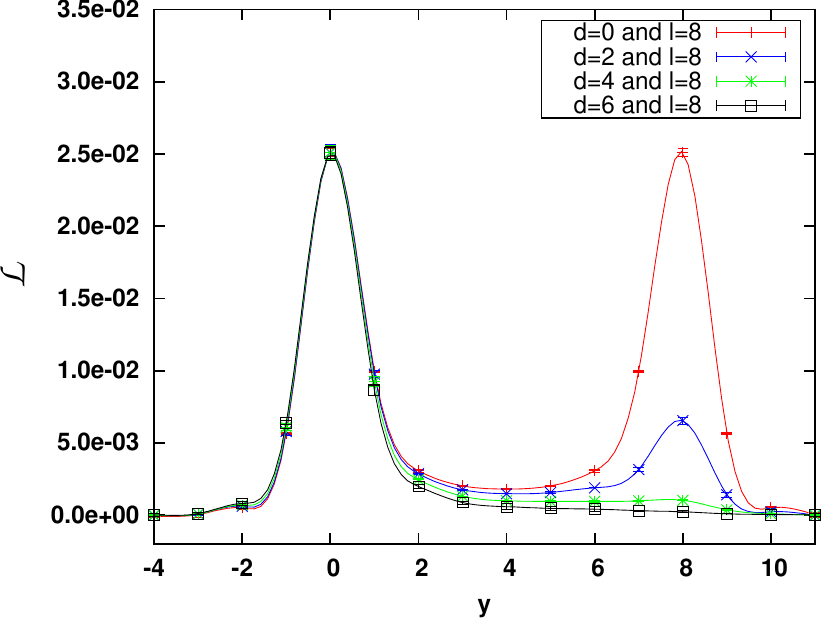}
\par\end{centering}}

    \subfloat[L geometry along the segment gluon-antiquark.\label{fig:action_L_gaq}]{
\begin{centering}
    \includegraphics[height=6cm]{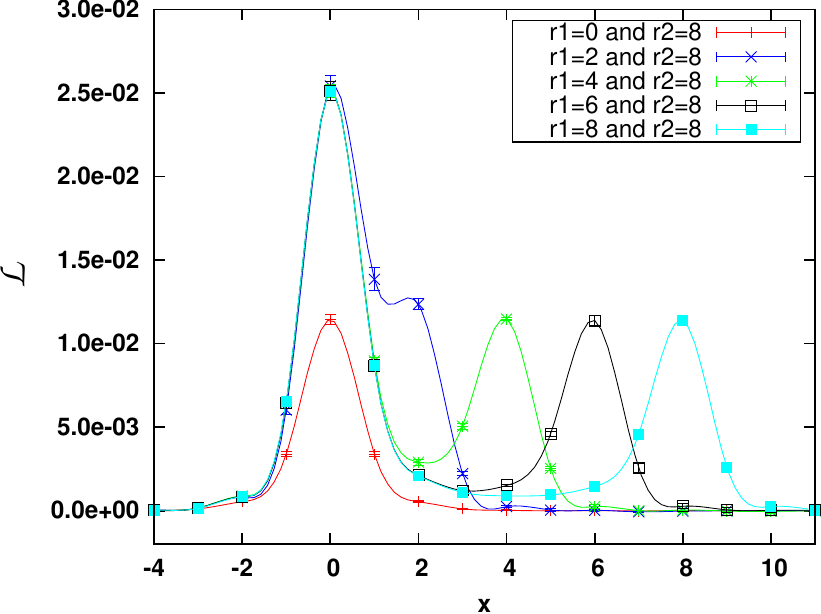}
\par\end{centering}}
    \subfloat[L geometry along the segment gluon-quark.\label{fig:action_L_gq}]{
\begin{centering}
    \includegraphics[height=6cm]{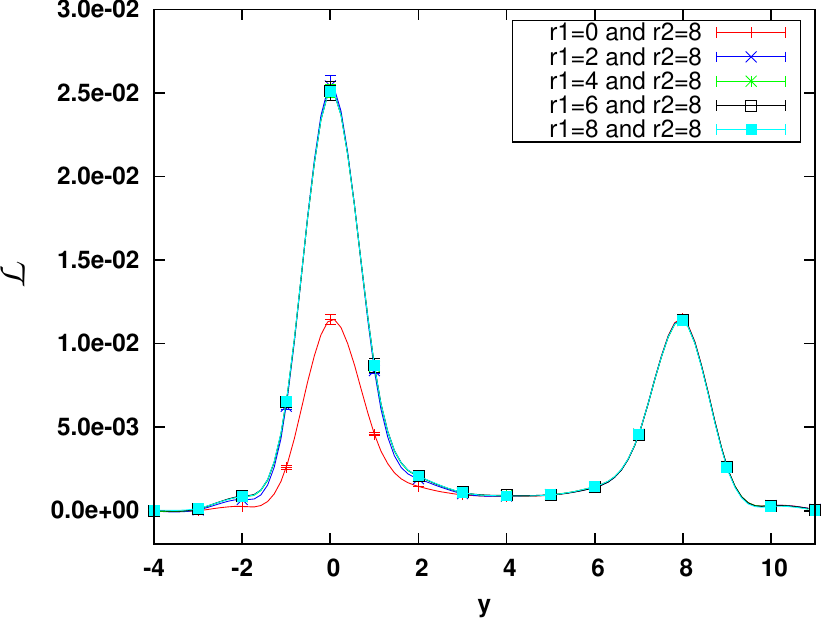}
\par\end{centering}}
\par\end{centering}
    \caption{Results for the lagrangian density. The lines were drawn for convenience and therefore do not represent results from any kind of interpolation. The results are in lattice spacing units.}
    \label{action_U_L}
\end{figure*}

\begin{figure*}
\begin{centering}
    \subfloat[$d=0$ and $l=8$\label{fig:cfield_qqg_U_Sim_d_0_l_8_Action}]{
\begin{centering}
    \includegraphics[height=4cm]{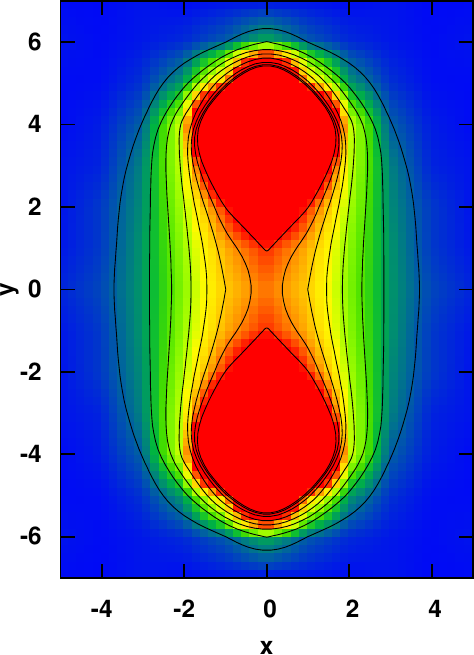}
\par\end{centering}}
    \subfloat[$d=2$ and $l=8$\label{fig:cfield_qqg_U_Sim_d_2_l_8_Action}]{
\begin{centering}
    \includegraphics[height=4cm]{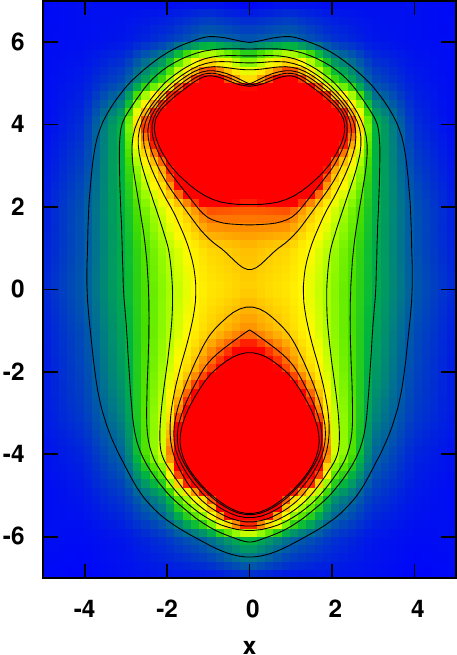}
\par\end{centering}}
    \subfloat[$d=4$ and $l=8$\label{fig:cfield_qqg_U_Sim_d_4_l_8_Action}]{
\begin{centering}
    \includegraphics[height=4cm]{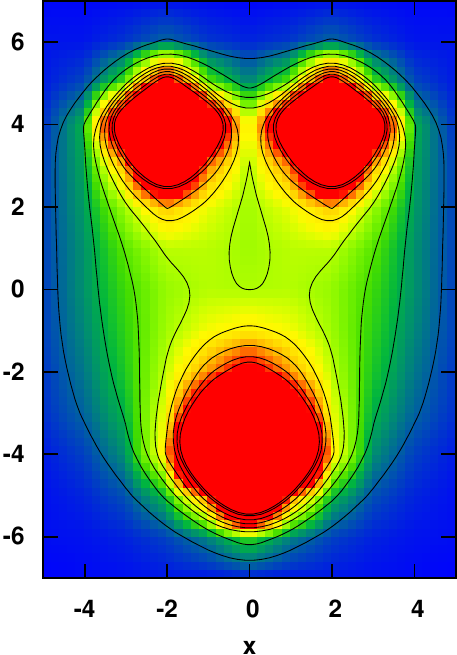}
\par\end{centering}}
    \subfloat[$d=6$ and $l=8$\label{fig:cfield_qqg_U_Sim_d_6_l_8_Action}]{
\begin{centering}
    \includegraphics[height=4cm]{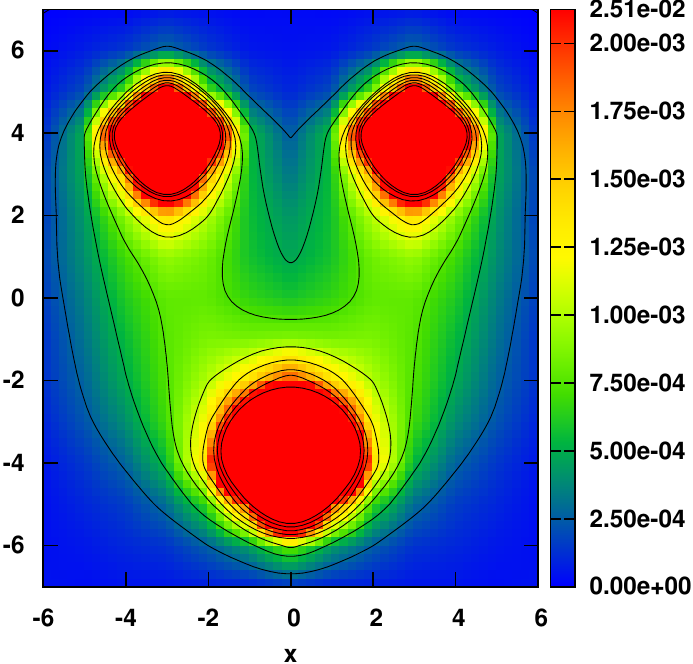}
\par\end{centering}}

    \subfloat[$r_1=2$ and $r_2=8$\label{fig:cfield_qqg_L_r1_2_r2_8_Action}]{
\begin{centering}
    \includegraphics[height=4cm]{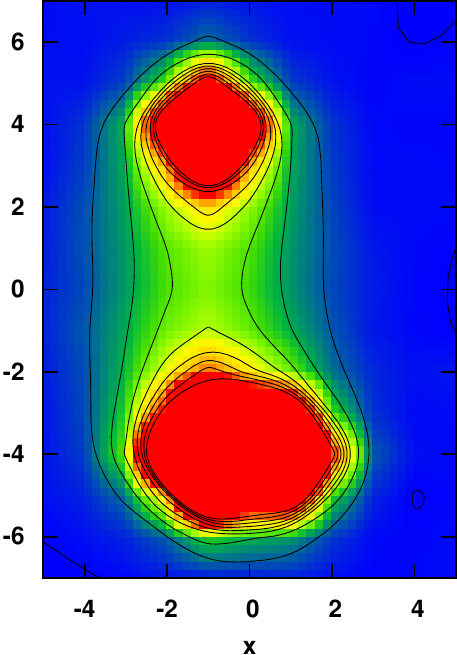}
\par\end{centering}}
    \subfloat[$r_1=4$ and $r_2=8$\label{fig:cfield_qqg_L_r1_4_r2_8_Action}]{
\begin{centering}
    \includegraphics[height=4cm]{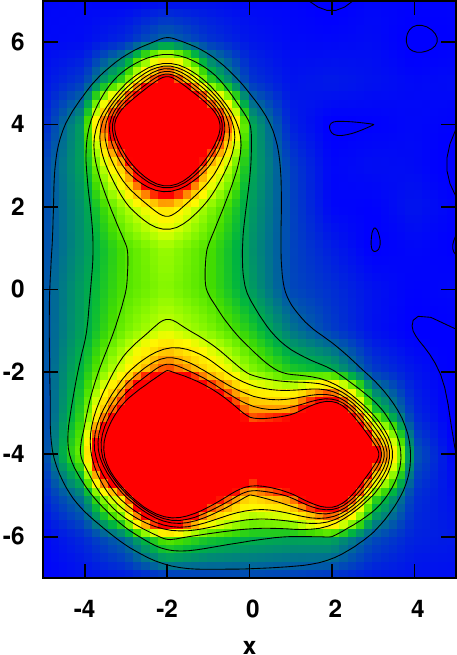}
\par\end{centering}}
    \subfloat[$r_1=6$ and $r_2=8$\label{fig:cfield_qqg_L_r1_6_r2_8_Action}]{
\begin{centering}
    \includegraphics[height=4cm]{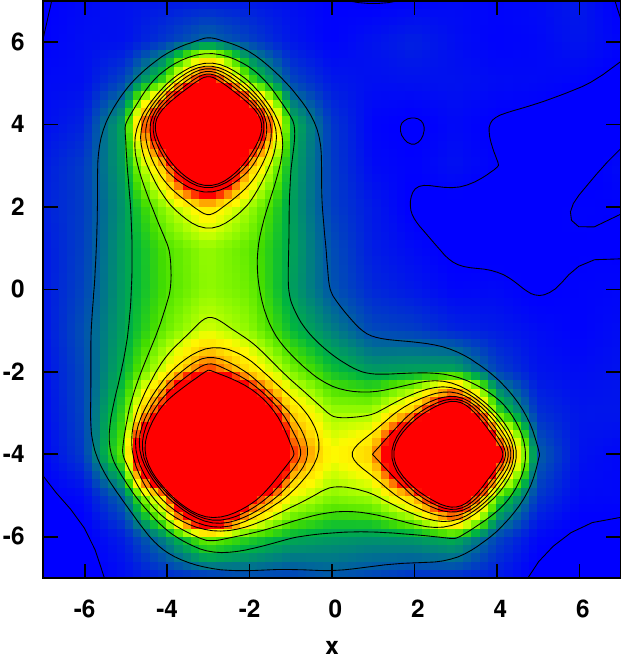}
\par\end{centering}}
    \subfloat[$r_1=8$ and $r_2=8$\label{fig:cfield_qqg_L_r1_8_r2_8_Action}]{
\begin{centering}
    \includegraphics[height=4cm]{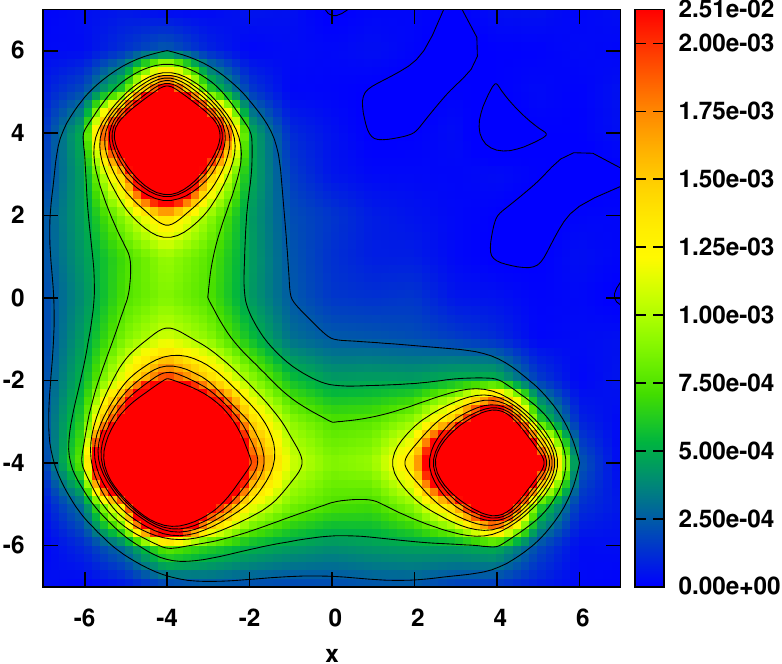}
\par\end{centering}}

\par\end{centering}
    \caption{Results for the lagrangian density. The Fig. \protect\subref{fig:cfield_qqg_U_Sim_d_0_l_8_Action}-\protect\subref{fig:cfield_qqg_U_Sim_d_6_l_8_Action} are for the U shape geometry and the Fig. \protect\subref{fig:cfield_qqg_L_r1_2_r2_8_Action}-\protect\subref{fig:cfield_qqg_L_r1_8_r2_8_Action} are for the L shape geometry. The top value of the colour scale is the maximum value of the field. The results are in lattice spacing units.}
    \label{cfield_qqg_U_L_Sim_Act}
\end{figure*}

\subsection{L Geometry}

The squared field components in the L geometry with $r_1 = r_2 = 8$  are shown in Fig.
\ref{cfield_qqg_L_r1_8_r2_8_Ei_Bi}.
In this figure, we can see that $\braket{E_x^2}$ is greater is the x axis and $\braket{E_y^2}$
is greater in the y-axis, on the other hand the chromomagnetic field components exhibit the reciprocal behaviour -
$|\braket{B_x^2}|$ is greater in the y axis and $|\braket{B_y^2}|$ is greater on the x axis. This result is consistent
with having two essentially independent fundamental strings, since this was the result obtained for one fundamental
string - the longitudinal component is the dominant one in the chromo-electric field and the transversal component is
dominant in the chromo-magnetic field.

In Fig. \ref{fig:cfield_qqg_L_r1_2_r2_8_Action}-\protect\subref*{fig:cfield_qqg_L_r1_8_r2_8_Action} , we show the distribution of the lagrangian density, in the L geometry, with
$r_2 = 8$, fixed, and for different $r_1$, where $r_1$ is the distance between gluon-antiquark and $r_2$ the distance between gluon-quark.
The variation of the lagrangian density with $r_1$ can also be seen in Fig.
\ref{fig:action_L_gaq} and in Fig. \ref{fig:action_L_gq}, in the $x$ and $y$ axis (Fig. \ref{shape}), where is the anti-quark and the quark.
Notice that the result in the $y$ axis is essentially the same, when we move the
antiquark in the $x$ axis, except for the case of $r_1 = 0$, where the system collapses in a meson.
But, even in this case, the flux tube near the quark is almost the same.

In the $y$ axis, we can see the presence of a flux tube between the gluon and the quark. As can be seen for
$r_2 = 8$, the lagrangian density tends to a constant in the center of the tube and remains practically unchanged when the antiquark and the gluon are far apart. This last result is consistent with the existence of a confining potential
$V_{gq} \to \sigma r$ between the gluon and the (anti)quark.

Our results indicate that in this geometry the system is well described by two independent fundamental strings
as was stated in \cite{Bicudo:2007xp} and \cite{Cardoso:2007dc}.

\subsection{U Geometry}

\begin{figure*}
\begin{centering}
    \subfloat[\label{fig:cfield_qqg_U_d_6_l_8_Ex}$\Braket{E_x^2}$]{
\begin{centering}
    \includegraphics[height=3.5cm]{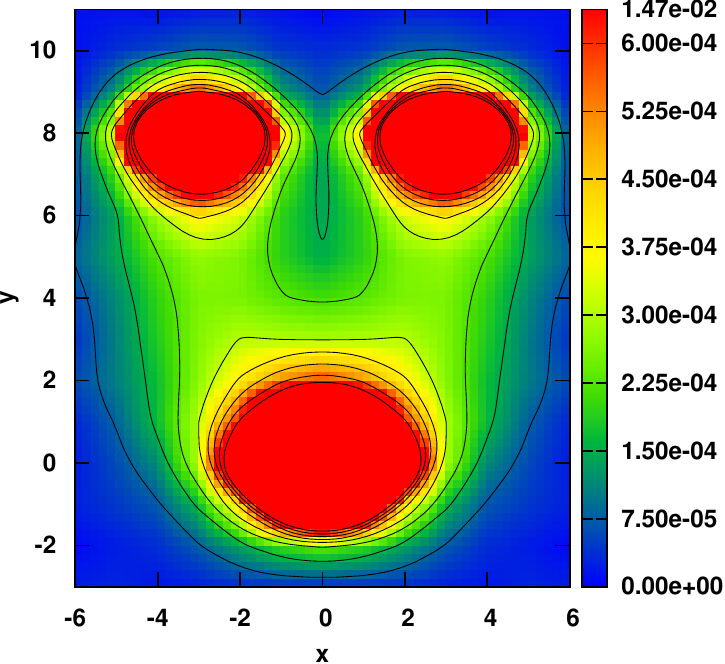}
\par\end{centering}}
    \subfloat[\label{fig:cfield_qqg_U_d_6_l_8_Ey}$\Braket{E_y^2}$]{
\begin{centering}
    \includegraphics[height=3.5cm]{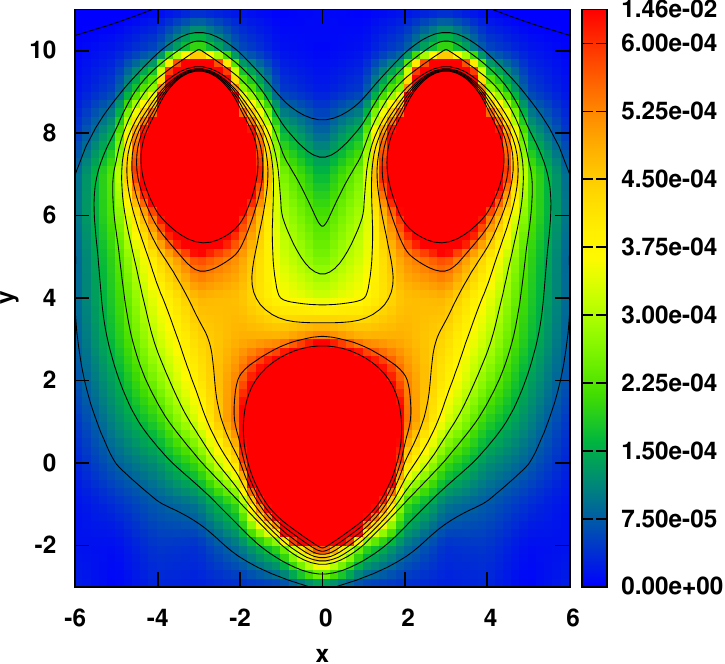}
\par\end{centering}}
    \subfloat[\label{fig:cfield_qqg_U_d_6_l_8_Ez}$\Braket{E_z^2}$]{
\begin{centering}
    \includegraphics[height=3.5cm]{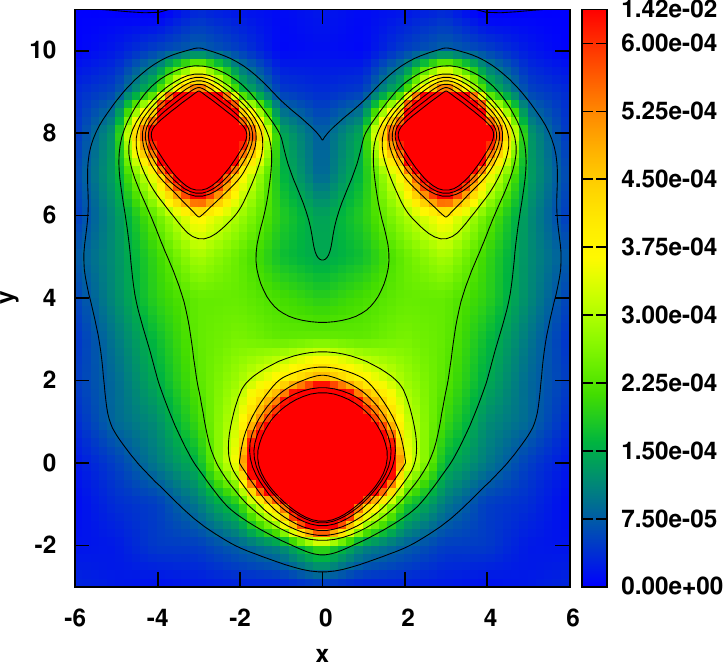}
\par\end{centering}}
    \subfloat[Energy Density\label{fig:cfield_qqg_U_d_6_l_8_Energ}]{
\begin{centering}
    \includegraphics[height=3.5cm]{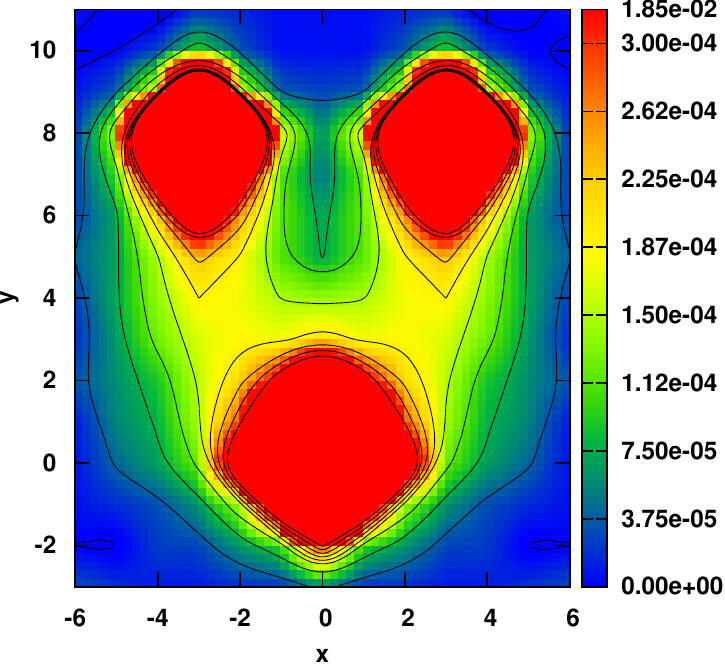}
\par\end{centering}}

    \subfloat[\label{fig:cfield_qqg_U_d_6_l_8_Bx}$-\Braket{B_x^2}$]{
\begin{centering}
    \includegraphics[height=3.5cm]{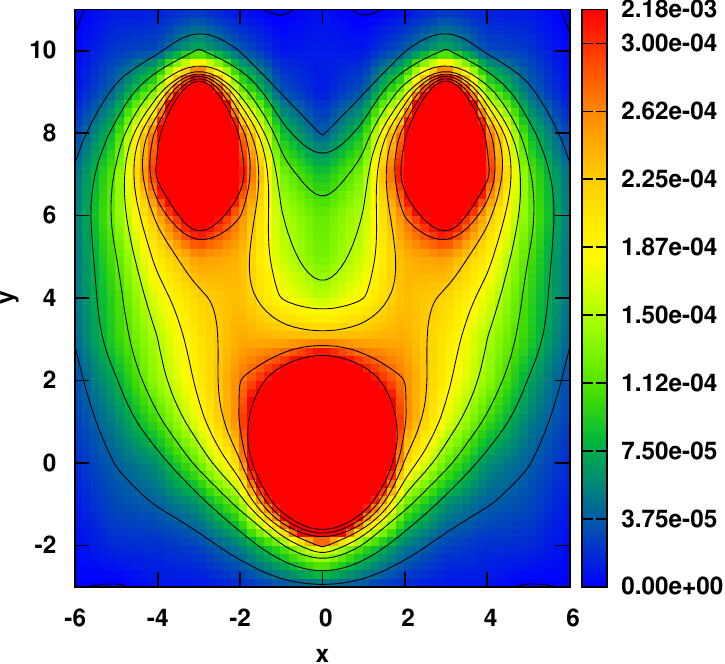}
\par\end{centering}}
    \subfloat[\label{fig:cfield_qqg_U_d_6_l_8_By}$-\Braket{B_y^2}$]{
\begin{centering}
    \includegraphics[height=3.5cm]{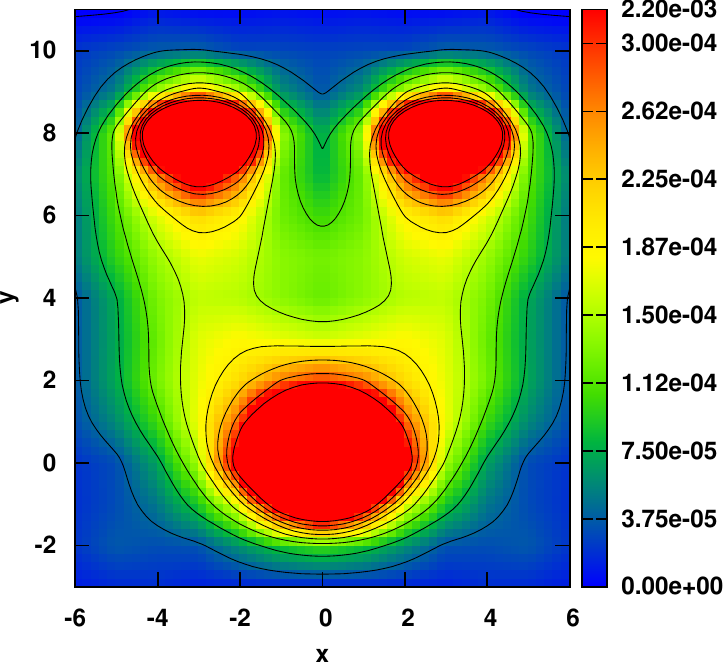}
\par\end{centering}}
    \subfloat[\label{fig:cfield_qqg_U_d_6_l_8_Bz}$-\Braket{B_z^2}$]{
\begin{centering}
    \includegraphics[height=3.5cm]{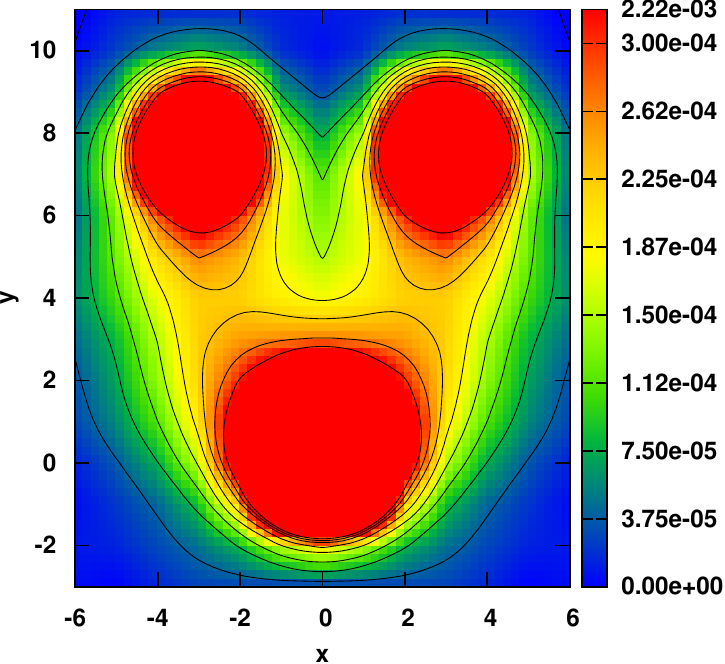}
\par\end{centering}}
    \subfloat[Lagrangian Density\label{fig:cfield_qqg_U_d_6_l_8_Act}]{
\begin{centering}
    \includegraphics[height=3.5cm]{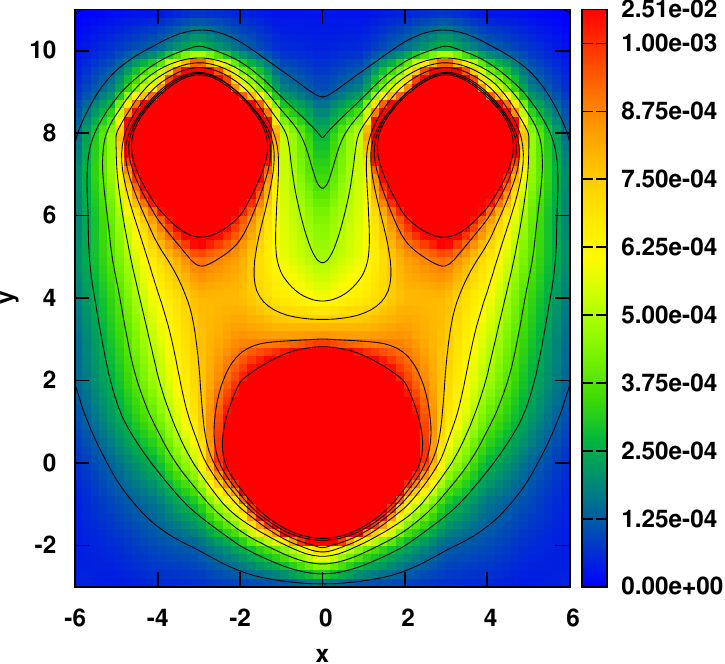}
\par\end{centering}}

\par\end{centering}
    \caption{Chromoelectric and chromomagnetic components and energy and lagrangian densities in the U shape geometry for $d=6$ and $l=8$. We use different colour scales to have a better view of the flux tube and the top value of the scale is the maximum value of the field. The results are in lattice spacing units.}
    \label{cfield_qqg_U_d_6_l_8_Ei_Bi}
\end{figure*}

We show the results for the chromoelectric and chromomagnetic fields in the U geometry at distances $l = 8$ and $d = 6$
in Fig. \ref{cfield_qqg_U_d_6_l_8_Ei_Bi}.
The results are consistent with the ones for the L shape geometry. The longitudinal component of the chromoelectric field is
the dominant component. This is the $y$ component of the chromoelectric, and this is expected since the flux tube is essentially
aligned in this direction. In the same way $\braket{B_x^2}$ and
$\braket{B_z^2}$ are seen to be dominant with relation to $\braket{B_y^2}$, which is consistent with the fact that the
transversal component of the magnetic is the larger one.

In Fig. \ref{fig:cfield_qqg_U_Sim_d_0_l_8_Action}-\protect\subref*{fig:cfield_qqg_U_Sim_d_6_l_8_Action} we can see the evolution of the lagrangian density, as a function of the
the quark-antiquark distance, $d$, for fixed $l = 8$. For $d = 0$, we are in the glueball case
and we thus have an adjoint string linking the two gluons. For $d = 2$ we can see the stretching of the tube in the
$x$ direction. For $d = 4$, corresponding to $y \simeq 5$, we already see the string splitting in two fundamental strings.
In d=6, the separation of the two fundamental strings is clear, they only join at the gluon position.
The transition point between the two regimes - one adjoint string and two clearly splitted fundamental ones - occurs
between $d=2$ and $d=4$, for $l=8$. This transition point occurs for an angle between the two fundamental strings
of $0.37 \pm 0.12 $ rad, and this is relevant for the quark and gluon constituent models.
In Fig. \ref{fig:action_U_g_qaq}, we can see the stretching and partial splitting of the flux tube in the equatorial plane
($y = 4$) between the quark and the antiquark, and in \ref{fig:action_U_gqq_x0} we see the results for the $y$ axis, where
the gluon is located (at $y = 0$), as well and the centroid of the $q\bar{q}$ subsystem (at $y = 8$).

\section{Conclusions}

We present the first study the chromoelectric and chromomagnetic fields produced by a static quark-gluon-antiquark system in a pure gauge SU(3) QCD lattice.

We report the cases of a simple meson and of a two gluon glueball, which correspond to two different degenerate cases of a hybrid meson system.
We verify the qualitative results for the squared components of the colour fields that were
obtained by other authors for the quark-antiquark system.
Namely, we find that the chromoelectric field is dominant over the chromomagnetic and that the longitudinal components of the chromoelectric field, as well as the transversal component of the chromomagnetic, are dominant over the other components of the respective fields.
We find a similar behaviour in the two gluon system.
We also verify that the results for the two degenerate systems are related, with the energy density
in the glueball flux tube being compatible with $9/4$ times the energy density in the meson flux tube.
This is in agreement with the Casimir scaling factor between the glueball and the meson, obtained by Bali \cite{Bali:2000un}.

We also study two geometries for the hybrid meson system. We study a L shaped geometry, with the gluon on the
origin, the quark on the $y$ axis and the antiquark on the $x$ axis.
In this case we verify the dominance of the
longitudinal component in the chromoelectric field and of the transversal component in the chromomagnetic field in the two
flux tubes coming from the gluon. We also concluded that this two flux tubes are, mainly, two independent fundamental
strings, which agrees with the results for the potential obtained by \cite{Bicudo:2007xp} and \cite{Cardoso:2007dc}.
We also study a U shaped geometry, which allow us to see the transition between the two regimes of confinement,
with one adjoint and with two splitted fundamental strings.

Whether the Casimir scaling is due to a repulsive superposition the two fundamental strings or to the actual existence
of an adjoint string, we cannot yet distinguish in the present study. We conjecture that both these
two pictures are essentially equivalent in the gluon-gluon system. But it appears that for angles between the
gluon-quark and gluon-antiquark segments larger than 0.4 rad, the two fundamental strings are splitted.
In the future, it will be interesting to complement the present study of the flux tubes,
with the computation of the static potential for the U geometry.

\acknowledgments
This work was financed by the FCT contracts POCI/FP/81933/2007 and CERN/FP/83582/2008.
We thank Orlando Oliveira for useful discussions and for sharing gauge field configurations..

\bibliographystyle{apsrev4-1}
\bibliography{paper}

\end{document}